\documentclass[aps,prl,two column]{revtex4-2}

\usepackage[linktocpage=true, colorlinks=true, urlcolor=blue, linkcolor=blue, citecolor=blue]{hyperref}
\usepackage{graphicx}
\usepackage{amsmath,amssymb}
\usepackage{color}
\usepackage{url}
\usepackage{hyperref}
\usepackage{cleveref}
\usepackage{xfrac}
\usepackage{upgreek}
\usepackage{siunitx}

\usepackage{lineno}
\usepackage[outdir=./]{epstopdf}

\usepackage{polski}
\usepackage[utf8]{inputenc}

\graphicspath{{./Images/}}

\begin{document}

\title{Diffuse-illumination holographic optical coherence tomography}

\author{L\'eo Puyo}
\affiliation{Institute of Biomedical Optics, University of L\"ubeck, Peter-Monnik-Weg 4, 23562 L\"ubeck, Germany}
\affiliation{gl.puyo@gmail.com}

\author{Clara Pf\"affle}
\affiliation{Institute of Biomedical Optics, University of L\"ubeck, Peter-Monnik-Weg 4, 23562 L\"ubeck, Germany}

\author{Hendrik Spahr}
\affiliation{Institute of Biomedical Optics, University of L\"ubeck, Peter-Monnik-Weg 4, 23562 L\"ubeck, Germany}

\author{Jonas Franke}
\affiliation{Institute of Biomedical Optics, University of L\"ubeck, Peter-Monnik-Weg 4, 23562 L\"ubeck, Germany}

\author{Daniel Bublitz}
\affiliation{Corporate Research \& Development, Zeiss AG, Jena, Germanyy}

\author{Dierck Hillmann}
\affiliation{Department of Physics and Astronomy, Vrije Universiteit Amsterdam, De Boelelaan 1081, 1081 HV Amsterdam, Netherlands}

\author{Gereon H\"uttmann}
\affiliation{Institute of Biomedical Optics, University of L\"ubeck, Peter-Monnik-Weg 4, 23562 L\"ubeck, Germany}
\affiliation{Medical Laser Center L\"ubeck GmbH, Peter-Monnik-Weg 4, 23562 L\"ubeck, Germany}
\affiliation{Airway Research Center North (ARCN), Member of the German Center for Lung Research (DZL), Wöhrendamm 80, 22927 Großhansdorf, Germany}


\begin{abstract}
Holographic optical coherence tomography (OCT) is a powerful imaging technique, but its ability to reveal low-reflectivity features is limited. In this study, we performed holographic OCT by incoherently averaging volumes with changing diffuse illumination of numerical aperture (NA) equal to the detection NA. While the reduction of speckle from singly scattered light is only modest, we discovered that speckle from multiply scattered light can be arbitrarily reduced, resulting in substantial improvements in image quality. This technique also offers the advantage of suppressing noises arising from spatial coherence, and can be implemented with a partially spatially incoherent light source for further mitigation of multiple scattering. Finally, we show that although holographic reconstruction capabilities are increasingly lost with decreasing spatial coherence, they can be retained over an axial range sufficient to standard OCT applications.
\end{abstract}

\maketitle

Full-field swept-source optical coherence tomography (FF-SS-OCT) employs a camera to record A-scans in parallel~\cite{Sarunic2006, Povazay2006, Dubey2007, Bonin2010}. FF-SS-OCT became holographic OCT with the ability to perform layer-wise digital refocusing of recorded volumes, effectively extending the available axial imaging range~\cite{Hillmann2011}. Holographic refocusing is made possible by the lateral phase stability intrinsic to full- or line-field detection. As digital holography also allows for the correction of higher order wavefront errors~\cite{Stadelmaier2000, Marron2009}, remarkably simple holographic OCT setups are able to perform diffraction limited imaging despite the presence of aberrations~\cite{Kumar2013, Hillmann2016, Ginner2017}. Holographic OCT has found its most impactful applications in ophthalmology, where its rapid volumetric imaging can freeze eye movements and reveal phase variations caused by nanometric axial expansion of the retinal tissue~\cite{Spahr2015, Hillmann2016PNAS}.

Multiple scattering remains a fundamental issue in holographic OCT. Multiply scattered light in OCT is detected at an incorrect depth due to the additional optical path length. Confocal gating is an effective strategy to reject multiply scattered light~\cite{Minsky1961}.
In point-scanning OCT, the tip of the single-mode fiber (SMF) plays the roles of the conjugate light source and pinhole. In Fourier-domain OCT, because the same optical focus is used for the entire A-scan, the confocal volume should be long enough to preserve resolution and sensitivity over most of the sample's thickness. Therefore, a tradeoff exists between axial imaging range and lateral resolution, often solved by using a lower than desired numerical aperture (NA). Thanks to the sensitivity and speed of Fourier-domain OCT~\cite{Leitgeb2003, ChomaSarunic2003, DeBoer2017}, modern commercial OCT systems used in ophthalmology are nevertheless based on swept-source or spectral-domain OCT. An NA of typically 0.05 allows most of the approximately $\SI{500}{\micro\metre}$ retinochoroidal thickness to fit in the Rayleigh range.
Such a low NA yields limited lateral resolution, but an often lesser considered consequence of the low NA is that it also hinders the effectiveness of multiple scattering rejection because multiply scattered light from the elongated confocal volume cannot be filtered~\cite{Leahy2016}.
Time-domain OCT is on the other hand not subject to this compromise between axial range and lateral resolution because the focus can be positioned precisely on the single layer being imaged. The NA can be increased and the focus dynamically adjusted to maintain the OCT voxel within the reduced confocal volume~\cite{Lexer1999, Qi2004dynamic, Dubois2018}. At high NA, the axial sectioning from the confocal gating can be made so short that it can even match or be shorter than the OCT temporal coherence gating~\cite{Dubois2004}. In this scenario, not only higher lateral resolution but also better multiple scattering rejection can be achieved with point- or line-scanning confocal time-domain OCT, allowing high-quality imaging in strongly diffusing samples such as the skin~\cite{Dubois2018}. As an alternative to physical pinholes that block out-of-focus light, coherence areas induced by limited spatial coherence also behave as virtual pinholes by preventing out-of-focus light from interfering, and by doing so, produce a form of confocality~\cite{Karamata2004, KaramataLaubscher2005_II, Barolle2021}. The illumination is said spatially incoherent when its spatial coherence matches or is smaller than the pixel size, as in incoherent time-domain full-field OCT~\cite{Beaurepaire1998, Dhalla2010}.
Alteration of the spatial coherence to reduce multiple scattering has also been demonstrated in holographic OCT with a fast deformable membrane~\cite{Stremplewski2019}, and a long multi-mode fiber (MMF)~\cite{Auksorius2022_MMFretina}. As a form of Fourier-domain OCT, holographic OCT must also ensure that confocal volumes stemming from coherence areas allow the detection of light backscattered from the entire sample's thickness, necessarily leading to an imperfect multiple scattering rejection. Unlike in scanning-OCT however, the size of the interferometric pinholes in holographic OCT can be decoupled from the pixel size. Coherence areas can be extended largely beyond the pixel size in order to elongate confocal volumes without sacrificing the imaging NA (i.e., the lateral resolution). Such a partial spatial incoherence was used to image the eye fundus with coherence areas larger than $\SI{150}{\micro\metre}$, allowing a multiple scattering rejection in the choroid similar to the one in commercial Fourier-domain scanning-OCT systems with a pixel resolution estimated to $\SI{4.6}{\micro\metre}$~\cite{Auksorius2022_iScience}.

Like other coherent imaging methods, holographic OCT is also susceptible to speckle-noise. Speckle-noise arises because light backscattered from different depths of voxels has different phase shifts, leading to destructive or constructive interference. The resulting sum of random complex phasors yields speckle grains that make the detected intensity in voxels differ from what is actually expected from their actual scattering coefficients~\cite{Dainty2013, Goodman2007speckle}. As illustrated from cases where speckle-noise could be cleared thanks to scatterers' motility, reducing speckle-noise is key to improving the visibility of low-reflectivity features in biological samples~\cite{Thouvenin2017, Liu2017imaging}.
Because the contribution of multiple scattering also manifests on OCT images as speckle, the two notions are sometimes confused. A distinction was however made as early as 1999 to distinguish signal-degrading and signal-carrying speckle, originating from multiply and singly scattered light, respectively~\cite{Schmitt1999}. The difference is important because these two sources of speckle have different characteristics and should be differently addressed~\cite{Stremplewski2019}. As speckle-noise originates from the coherence of light within one imaging element, reducing the spatial coherence -even to the size of a pixel- does not reduce speckle-noise.
Speckle-noise can admittedly be reduced digitally, typically with more or less sophisticated forms of spatial averaging~\cite{Ozcan2007, Bianco2018, See1994, Argenti2013}, but such filtering leads to a loss of spatial resolution. Another approach to reduce speckle-noise is to incoherently average images with independent speckles. Provided the $M$ averaged speckles are uncorrelated, speckle-noise is reduced by a factor of $\sqrt{M}$~\cite{Goodman2007speckle}. Independent speckles are typically obtained by modifying the illumination or detection angle, or the wavelength and polarization~\cite{Szkulmowski2012}.
However, obtaining a sufficient number of independent speckles is challenging. Winetraub et al. investigated the potential of angular compounding in scanning Fourier-domain OCT to reduce speckle-noise~\cite{Winetraub2019}. They found that manipulation of the illumination with angular compounding over a full illumination NA of $13^\circ$ could only yield a maximum speckle-noise reduction by a factor of 1.3, and that a better reduction necessarily implies loss of spatial resolution. They concluded that when a greater reduction is observed, other effects must be contributing, such as spatial averaging. Similarly, Desjardins et al. found that the achievable speckle-noise reduction is limited by the illumination NA~\cite{Desjardins2007}. Stremplewski et al. performed angular compounding in Fourier-domain full-field OCT and found that speckle-noise could only be reduced by a factor of less than 2~\cite{Stremplewski2019}. The rationale behind this shared conclusion is that greater illumination angles are better able to induce larger phase difference between scatterers, and therefore to obtain a greater number of independent speckles.

In 2017, Liba et al. performed spectral-domain OCT imaging with a rotating diffuser to illuminate the sample with a modulated speckle-field~\cite{Liba2017}. They showed that although the use of the diffuser slightly worsens individual OCT volumes, the reduction of speckle-noise when averaging multiple frames leads to significantly improved OCT images. Our work, also inspired by a patent~\cite{Bublitz2021}, aimed at adapting this approach to holographic OCT. We performed holographic OCT with a slowly changing diffuse illumination and incoherent averaging of several volumes, and we studied the reduction of speckle, multiple scattering, and of noises arising from spatial coherence. To further reduce the influence of multiple scattering, we combined the diffuse illumination to the use of a long MMF demonstrated by Auksorius et al.~\cite{Auksorius2022_MMFretina}. Finally, we investigated the limitations on the axial imaging range that come with the reduction of spatial coherence and the high illumination NA.


\section{Methods}

\begin{figure}[t!]
\centering
\includegraphics[width = 1\linewidth]{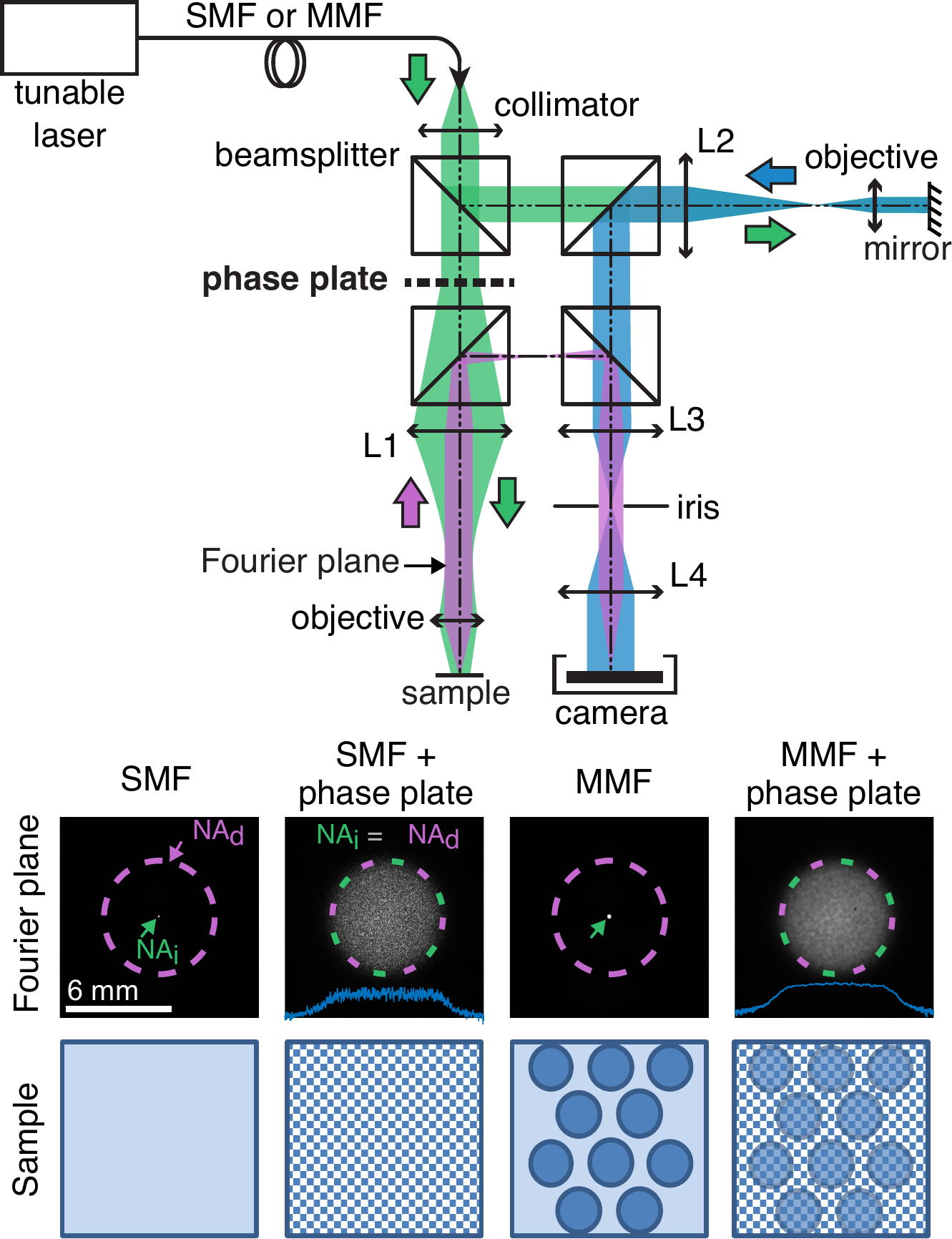}
\caption{Holographic OCT setup. SMF and MMF: single- and multi-mode fiber. L1-L4: converging lenses. $\rm{NA}_{\rm{i}}$ and $\rm{NA}_{\rm{d}}$: illumination and detection NA, respectively. The phase plate is placed in a plane conjugate to the sample, and diffuses the illumination to ensure $\rm{NA}_{\rm{i}} = \rm{NA}_{\rm{d}}$. In the sample plane, the MMF structures the illumination field with coherence areas whereas the phase plate projects an illumination pattern with high-spatial frequencies.
}
\label{1_Setup_Streuwellen_OAMZ}
\end{figure}

We conducted our experiments using the Mach-Zehnder interferometer depicted in Fig.~\ref{1_Setup_Streuwellen_OAMZ}, with a swept-source laser diode (Superlum BroadSweeper BS-840-1) offering a $\SI{75}{\nano\meter}$ tuning range centered around $\lambda_0 = \SI{840}{\nano\meter}$ and $\SI{3}{\milli\watt}$ of output power. The laser is coupled to either a single- or multi-mode fiber (SMF and MMF). The MMF has a core diameter of $\SI{50}{\micro\metre}$ and is $\SI{300}{\metre}$ long (Thorlabs FG050LGA, low OH, 0.22 NA). The extended length ensures that all modes emerging from the MMF are incoherent with each other due to the laser's limited temporal coherence~\cite{Kim2005, Dhalla2010, Auksorius2021_MMFcornea}. Mutual incoherence between the excited MMF modes results in the formation of coherence areas at the fiber tip. The number of incoherent modes depend on the fiber's length, diameter and NA, and on the laser's wavelength and temporal coherence~\cite{Auksorius2021_MMFcornea, Auksorius2022_MMFretina}. By conservation of the number of modes, the same number of coherence areas is found in the sample plane and fiber tip~\cite{BornWolf1993}. With the same MMF and a very similar optical configuration, Auksorius et al. estimated that the MMF produced approximately 250 incoherent modes~\cite{Auksorius2022_MMFretina}.
The fiber is inserted into a collimator with a $\SI{36}{\milli\metre}$ focal length, and the emerging beam is split into two arms, each containing a lens (L1, L2) with a focal length of $\SI{150}{\milli\metre}$ and a microscope objective (Plan Achromat, $10\times$, 0.25 NA, Olympus). Light from the sample and mirror is recombined with a fourth beamsplitter and relayed onto the camera with a pair of lenses with $\SI{100}{\milli\metre}$ focal length. The sample to camera magnification factor is $M=8.3$.
Frames of size 384 $\times$ 256 pixels are recorded at $\SI{60}{\kilo\hertz}$ with a high-speed CMOS camera (Photron FASTCAM SA-Z). The iris diaphragm between L3 and L4 filters light from the sample to avoid having spatially undersampled interference fringes on the camera. The iris diameter $D$ is therefore set to limit the bandwidth of the interference signal $K_{\rm O} =  \frac{\pi \cdot D}{\lambda_0 \cdot f} $, where $f$ is the focal length of L4, to make it fit into the camera's Nyquist bandwidth $K = \frac{\pi}{\Delta x}$, where $\Delta x = \SI{20}{\micro\metre}$ is the camera pixel size. This imposes an iris diameter of $\SI{4}{\milli\metre}$, which is magnified into a virtual diaphragm of $\SI{6}{\milli\metre}$ in the Fourier plane of the objective. The detection NA is calculated as ${\rm NA_{d}}= M \lambda_0 / 2 \Delta x$. We have here ${\rm NA_{d}}= 0.174$, which fits in the 0.25 objective NA.

The phase plate, or scattering disc, is a holographic diffuser that was designed and fabricated by photolithography for this experiment by the company Zeiss.
It was designed to have a $1.1^\circ$ scattering angle which suits the needs of the setup, and to give a top-hat intensity distribution in the Fourier plane.
This phase plate is positioned in the focal plane of L1 and is conjugate to the sample and camera with a magnification of 1:8.3 and 1:1, respectively. The effect of the phase plate has to be considered in both the Fourier and object plane.
As a scattering disc, it diffuses light to produce a uniform top-hat intensity distribution over a circular area of $\SI{6}{\milli\metre}$ of diameter in the pupil plane of the objective. This exactly matches the extent of the virtual detection aperture in that plane, meaning the illumination NA is equal to the detection NA.
The Fourier plane images in Fig.~\ref{1_Setup_Streuwellen_OAMZ} were obtained by placing a camera in the pupil plane of the objective to record the distribution of intensity for a single wavelength.
Without scattering disc, the $\SI{5}{\micro\metre}$ SMF or $\SI{50}{\micro\metre}$ MMF tip is magnified by 4.2 over an area with a diameter of approximately $\SI{30}{\micro\metre}$ or $\SI{200}{\micro\metre}$, respectively. The scattering disc spreads this energy over an area with a diameter of $\SI{6}{\milli\metre}$, thereby reducing the surface irradiance by a factor of 40,000 and 900 for the SMF and the MMF, respectively.
The sample plane schematics represent how the illumination field is structured by the MMF and the phase plate. With SMF, the sample is uniformly illuminated with Gaussian beam effects, whereas the MMF creates spatial coherence areas that are much larger than pixels~\cite{Auksorius2022_MMFretina}. Due to the illumination NA being equal to the detection NA, the phase plate projects an illumination pattern with spatial frequencies extending up to the camera's Nyquist limit.
Combining the MMF to the phase plate multiplexes those two ways of structuring the illumination field. It is important to note that although the phase plate reduces the spatial autocorrelation of the electric field, it does however not alter the spatial coherence in the sample plane. Spatial coherence, i.e., the extent of so-called coherence areas over which the temporal variations (ability of light to interfere) of the electric field are correlated, remains determined by the number of incoherent modes emerging from the MMF.

The phase plate is manually moved during measurements, slowly enough that it can be considered static during the acquisition of one volume ($\SI{8.3}{\milli\second}$), but fast enough that speckles are changed from one volume to the next ($\approx \SI{150}{\milli\second}$). This stands in contrast with methods where the light field is changed faster than the $\approx \SI{15}{\micro\second}$ camera exposure time for the purpose of reducing the spatial areas of temporal coherence, e.g., with a fast deformable membrane~\cite{Stremplewski2019}.
Complex holograms were high-pass filtered to remove large speckles resulting from residual interference between incoherent modes of the MMF~\cite{Auksorius2021_MMFcornea}.
OCT volumes were corrected for defocus with a phase factor determined by visual assessment of the images resolution, and typically 100 volumes were averaged incoherently.

\section{Speckle \& multiple scattering reduction}

\begin{figure}[t!]
\centering
\includegraphics[width = 1\linewidth]{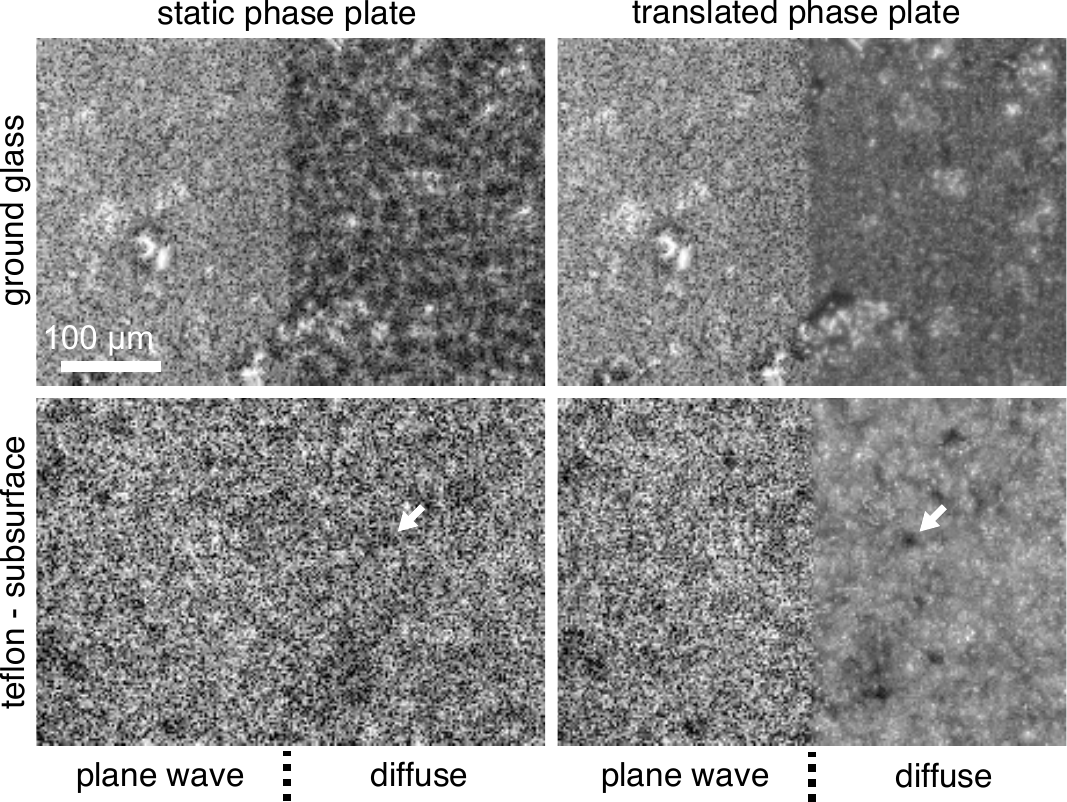}
\caption{Holographic OCT \textit{en-face} images with static and translated diffuse illumination. Only the right half of the field of view was diffusely illuminated. In single scattering conditions such as with ground glass, the illumination speckle-field deteriorates the quality of individual volumes. In a diffusing sample such as teflon, the illumination pattern is not seen. In both cases, significant speckle-reduction is observed when translating the phase plate during the measurement of the 100 incoherently averaged volumes.
}
\label{2_PPP_examples}
\end{figure}

The effects of diffuse illumination are first shown in Fig.~\ref{2_PPP_examples} with images of ground glass (Thorlabs, DG10-220) and a volume scatterer made of teflon. The phase plate was only partially positioned on the optical path of the illumination beam so that only the right half of the field of view was diffusely illuminated.
These measurements were made with the MMF.
On the left and right are shown images recorded with a static and vertically translated phase plate, respectively.
When imaging ground glass and keeping the phase plate static, the diffuse illumination pattern is very visible, which deteriorates the OCT imaging quality. The illumination pattern is however not visible when imaging the teflon sample below the surface: illumination with diffuse light or plane wave cannot be distinguished. The difference of visibility of illumination pattern between the two samples can be explained by the difference in scattering properties of the detected light. Ground glass is a one-layer sample from which we expect to detect mostly singly scattered light. In this case, the illumination pattern should be well visible. The fact that the projected speckle pattern is not visible in teflon indicates that the illuminating light got diffused within the sample, and therefore that multiply scattered light contributes to the signal detected from this layer. Multiple scattering is expected since teflon is imaged below the surface.
Translating the phase plate during the measurement of the 100 averaged volumes randomly changes the angular distribution of diffuse illumination of each voxel from one volume to the next. This changes the phase differences between scatterers and therefore leads to different speckle realizations in all volumes. A speckle reduction due to the diffuse illumination is clearly visible in both samples when comparing with the part of the field of view illuminated with a plane wave. The quality of OCT images is significantly improved as the sample features of low reflectivity become much better visible (arrows). For ground glass, part of the improvement also comes from the averaging of the illumination intensity pattern.

\begin{figure}[t!]
\centering
\includegraphics[width = 1\linewidth]{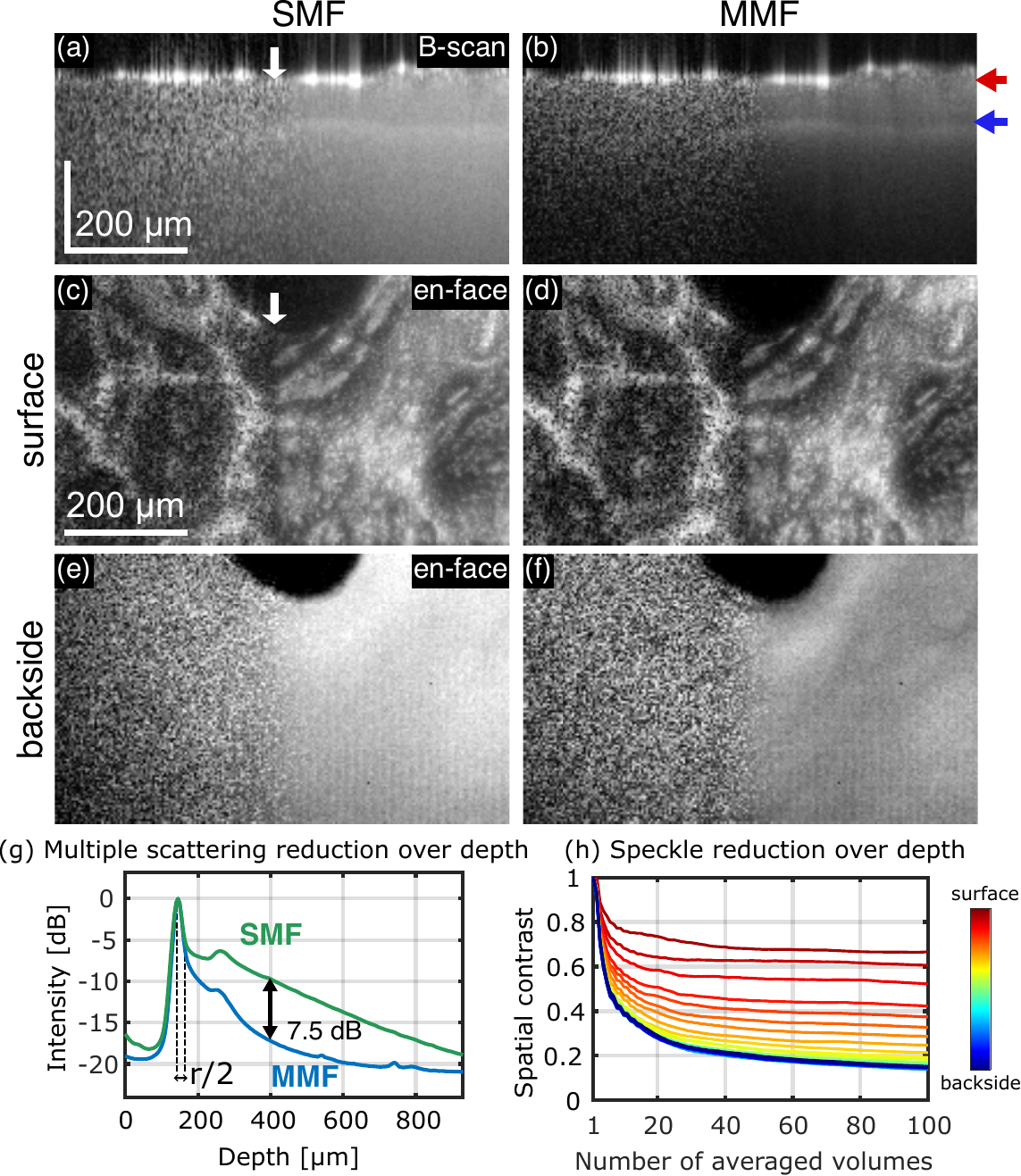}
\caption{Speckle-noise and multiple scattering reduction by combining phase plate and a MMF. A rubber glove is imaged using a SMF and a MMF and with only half of the field of view diffusely illuminated.
(a-b) B-scans. 
(c-f) \textit{en-face} image of the sample's surface and (e-f) deepest layer.
(g) Intensity profiles of B-scans with SMF and MMF.
(h) Local spatial contrast calculated at all sample's depth and for all number of averaged volumes.
}
\label{3_DepthSample}
\end{figure}

Images of a rubber glove imaged with a SMF and a MMF with again only half on the sample diffusely illuminated are shown in Fig.~\ref{3_DepthSample}. The phase plate was translated during the measurement and 100 volumes were incoherently averaged. For each measurement are shown one B-scan and one \textit{en-face} image of the surface and backside of the glove (see arrows on B-scan).
It should be noted that the visibility of the backside of such type of glove varies according to their properties, e.g., thickness.
Combining the MMF to the diffuse illumination yields images where both multiple scattering and speckle-noise are reduced. The effects of the reduced spatial coherence and of the diffuse illumination can be separately considered.

The comparison of cross-sectional views using a SMF or a MMF shows how the partial spatial incoherence reduces the intensity of multiply scattered light detected below the sample. A multiple scattering reduction of $\SI{7.5}{\decibel}$ between the SMF and the MMF is measured in Fig.~\ref{3_DepthSample}(g) by comparing the normalized intensity profiles over depth for the two measurements.
This reduction in multiple scattering is related to the number of incoherent modes~\cite{Auksorius2022_MMFretina}, but it is important to note that the reduction is not the same in all layers.
The radius $r$ of coherence areas can be estimated by comparing the intensity profiles over depth obtained using the SMF and the MMF with the following reasoning. The shortest distance beyond which light must be laterally scattered to be prevented from interfering is the radius of coherence areas. Light laterally scattered at distance $r$ and scattered again to the camera will be detected by OCT at a depth below the actual scattering layer by $r/2$. Therefore, the MMF only reduces multiple scattering in layers that are $r/2$ below the diffusing layer. By comparing the depth profiles measured with the SMF and the MMF, the distance between the surface and the depth from which multiple scattering starts to be rejected can be estimated to approximately $\SI{20}{\micro\metre}$. This gives a value of $\SI{80}{\micro\metre}$ for the diameter of coherence areas.

As seen before, the phase plate on the other hand reduces speckle. Although significant speckle reduction is visible on the entire B-scan, a notable difference is observed among layers: the effect is important at depth but only modest on the surface. The efficiency of speckle reduction over depth is investigated in Fig.~\ref{3_DepthSample}(h).
For each layer and for each number of averaged volumes, the local spatial contrast (standard deviation divided by average intensity) was computed on small regions of interest spanning over the diffusely illuminated field of view.
The reduction in local spatial contrast reveals the reduction in speckle-noise. The sample's reflectivity must however be sufficiently uniform within the local window so that the spatial variations of intensity are due to speckle and not to actual differences in intensity from the sample. The graphs show that only little improvement is achieved in the surface layers as spatial contrast is only reduced from 1 to about 0.6. On the other hand, the speckle reduction observed in the deepest layers almost follows the law of independent speckles, i.e., $1/ \sqrt{100}=0.1$. As the saturation in spatial contrast reduction may be due to either residual speckle-noise or existing spatial features, it is not possible to say here whether speckle could be further reduced.
However, given the reports mentioned in the Introduction about speckle-noise reduction with angular compounding, a factor of almost 10 in speckle reduction as observed in the sample's deepest layers is already much more than expected. This is because this limit in achievable speckle-reduction by angular compounding concerns singly scattered light and not multiply scattered light. However, as visible on B-scans, the signal detected from the layers in depth is mainly constituted of multiply scattered light. The air/glove (red arrow) and glove/air interfaces (blue arrow) corresponding to the glove's surface and backside should give a similar specular reflection. The absence of specular reflection at the backside interface indicates that the signal from this depth has been diffused. The prominence of multiply scattered light detected at greater depth is a known phenomenon in OCT~\cite{Hillman2006}, and is due to the limited capabilities of Fourier-domain OCT to reject multiple scattering. The presence of remaining multiple scattering can be expected because the partial spatial incoherence fails to reject light multiply scattered within coherence areas, and coherence areas are $\SI{80}{\micro\metre}$ large, which is much larger than the $\SI{2.4}{\micro\metre}$ pixel size.


It is important to note that the diffuse illumination and the MMF reduce the influence of multiply scattered light in different ways. The partial spatial incoherence reduces the intensity of multiple scattering by preventing it from being detected. The diffuse illumination suppresses the speckle-noise of the detected multiply scattered light. This second way is arguably not as powerful, but it nonetheless largely improves images. The smoothed multiply scattered light signal can be subtracted, and the diffuse illumination can reduce the influence of multiple scattering within coherence areas, i.e., within the sample's surface layers. In this example, the glove/air interface (blue arrow) is better visible with the SMF and phase plate than with the MMF without phase plate.

\section{Spatial coherence noise \& lateral resolution}

\begin{figure}[t!]
\centering
\includegraphics[width = 1\linewidth]{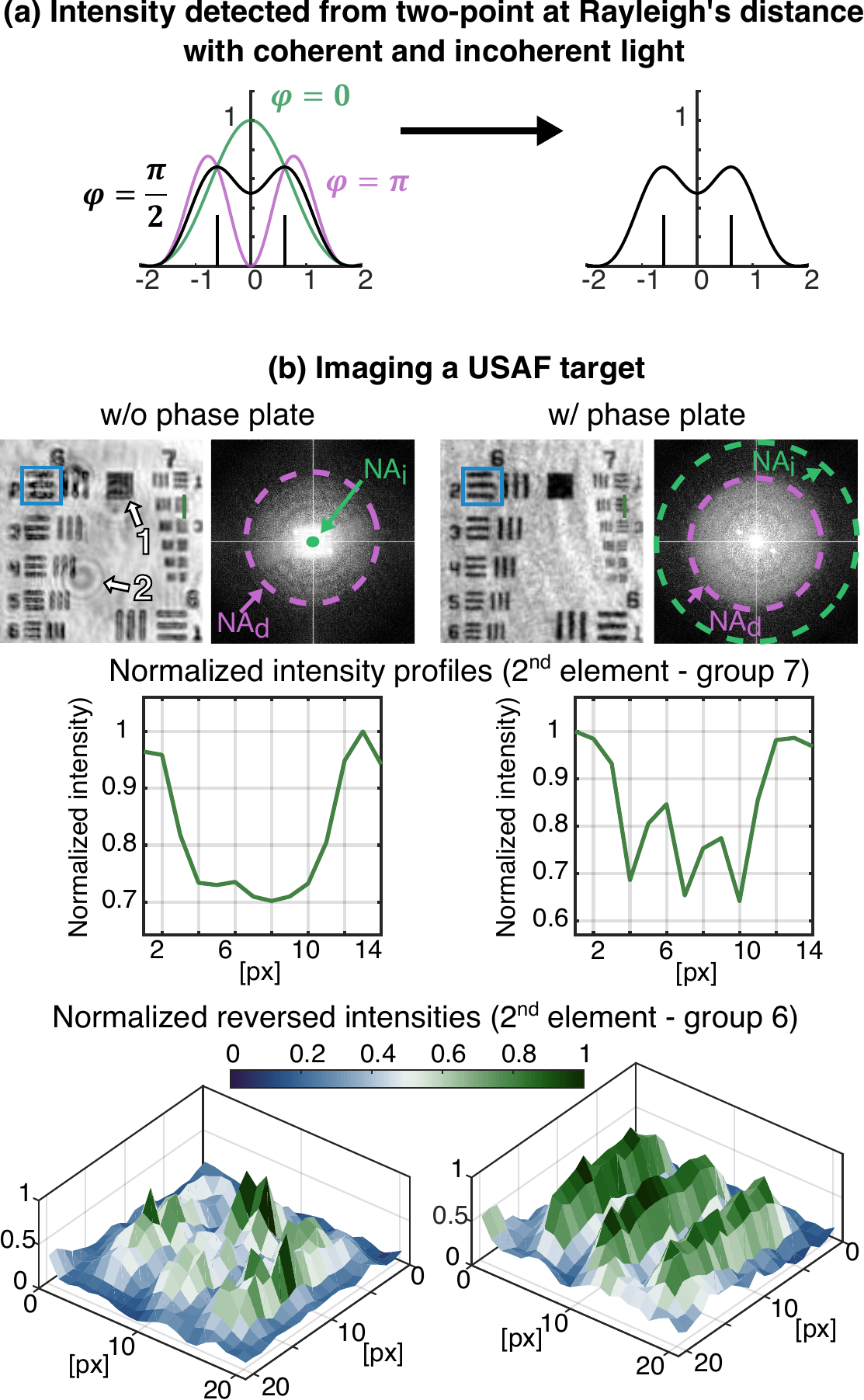}
\caption{Suppression of spatial coherence-noises with diffuse illumination.
(a) With coherent light, the resolution of two-points at Rayleigh's distance depends on their phase difference.
(b) A USAF target is imaged without and with phase plate. The illumination and detection NAs used are indicated on the Fourier transform of the complex holograms. With phase plate, the contrast of large structures and the resolution of smaller features are improved.
}
\label{5_USAF_Target_IncreasedResolution}
\end{figure}

Diffuse illumination can also improve OCT images by reducing noises arising from spatial coherence effects occurring at a scale much smaller than the size of coherence areas, e.g., unwanted interference between light scattered by adjacent structures. This type of coherent noises stands in contrast to speckle-noise, as the latter stems from the interference effects within voxels. Coherence noise particularly influences the resolution of two-points at Rayleigh's resolution distance. In the absence of aberration, light from the scatterers focuses over the Airy disc point-spread function. If the light field illuminating the scatterers is coherent, the complex fields from the scatterers add up so that the detected intensity is $I_{\rm{coherent}} =  {E_1}^2 +  {E_2}^2 + 2 E_1 E_2 \operatorname{Re}(e^{i\varphi})$, where $E_1$ and $E_2$ are the magnitudes of the electric fields emitted by the scatterers and $\varphi$ is their phase difference. As illustrated in Fig.~\ref{5_USAF_Target_IncreasedResolution}(a), the two points are resolved only if $\varphi$ is favorable. With incoherent light, the intensities of the electric fields add up so that the detected intensity is $I_{\rm{incoherent}}=  {E_1}^2 + {E_2}^2 $, and the two points are resolved independently of their phase difference. Overall, spatial coherence noises degrade the imaging quality by making the detected OCT intensity sensitive to local phase differences. When restricting the spatial coherence to the pixel size by filling the illumination pupil in order to have $\rm{NA_{i}} = \rm{NA_{d}}$, the light fields emitted by adjacent pixels become incoherent with one another and coherence-noises are suppressed. In holographic OCT, since the light is spatially coherent, when using diffuse illumination and averaging incoherently $N$ volumes with different phase plate states, the resulting intensity is:
\begin{equation}
\langle {I}_{\rm coherent} \rangle_N =  {E_1}^2  +  {E_2}^2
+ 2 \dfrac{ E_1 E_2}{N} \sum_{n=1}^{N} \operatorname{Re}(e^{i\varphi_n})
\end{equation}
When the phases $\varphi_n$ are uncorrelated and the number of averaged volumes $N$ is sufficiently large, the sum of complex phasors becomes negligible compared to the term ${E_1}^2 + {E_2}^2$. In that case, the detected intensity becomes the same as when imaging with incoherent light. Since the illumination NA extends to Nyquist's limit, the phase plate is able to project a pattern supporting frequencies great enough to illuminate adjacent pixels with different phases. It is however challenging to prove that the distribution of phase would be in practice sufficiently randomly distributed to suppress the average interference term.
The effects of diffuse illumination on spatial coherence noises are therefore verified experimentally by imaging a negative USAF target without and with the rotating phase plate, i.e., with plane wave and diffuse illumination, respectively. The detection NA was decreased using the instrument's iris to better visualize the effect due to the larger point-spread function. The NA of the diffuse illumination is therefore slightly greater than the detection NA.
These measurements were made with the MMF.
The two OCT images obtained without and with phase plate are shown in Fig.~\ref{5_USAF_Target_IncreasedResolution}(b), as well as the Fourier transform of the corresponding complex holograms. The purple dashed circle shows the extent of the detection NA. Without phase plate, the illumination NA is very small (green arrow), whereas with phase plate the illumination NA (green dashed circle) still matches the accessible camera bandwidth. In the plots below are shown the normalized intensity profiles for the second element of group 7. These are resolved only with the diffuse illumination and incoherent averaging. At the bottom of Fig.~\ref{5_USAF_Target_IncreasedResolution}(b) is shown the reversed normalized intensity for the second element of group 6. Although these features are much larger than diffraction limit, they are nonetheless much better imaged when using the phase plate. Similarly, the uniformly reflecting black square (arrow 1) can be imaged correctly with diffuse illumination. Other notable effects of the suppression of spatial coherence noises is the removal of diffraction rings (arrow 2) typically caused by dust on lenses~\cite{Park2009}, and the suppression of ringing artifacts, which allows a smoother imaging of sharp edges~\cite{Goodman2005}.

The minimally resolvable distance is reduced thanks to the diffuse illumination, but this arguably does not constitute an improvement in resolution since the detected highest spatial frequency of the sample is not increased. The diffuse illumination allows obtaining images with the advantageous incoherent contrast, and as detailed by Goodman, the difference between coherent and incoherent imaging is more a matter of contrast than of resolution~\cite{Goodman2005}. Coherent noise suppression is particularly useful for a sample such as a USAF target. The variations of surface between adjacent pixels of this object are very flat relatively to the wavelength, so that adjacent pixels have the same phase. This is the most unfavorable case to resolve two points with coherent light. A flat surface also causes dusts or imperfections to deteriorate the contrast. For biological samples, the axial size of OCT voxel largely exceeds the wavelength and the axial distribution of scatterers is random. The phase between adjacent pixels is therefore random and no significant improvement in resolving power is obtained through coherence-noise reduction.

\section{Spatial coherence \& axial range}

\begin{figure}[t!]
\centering
\includegraphics[width = 1\linewidth]{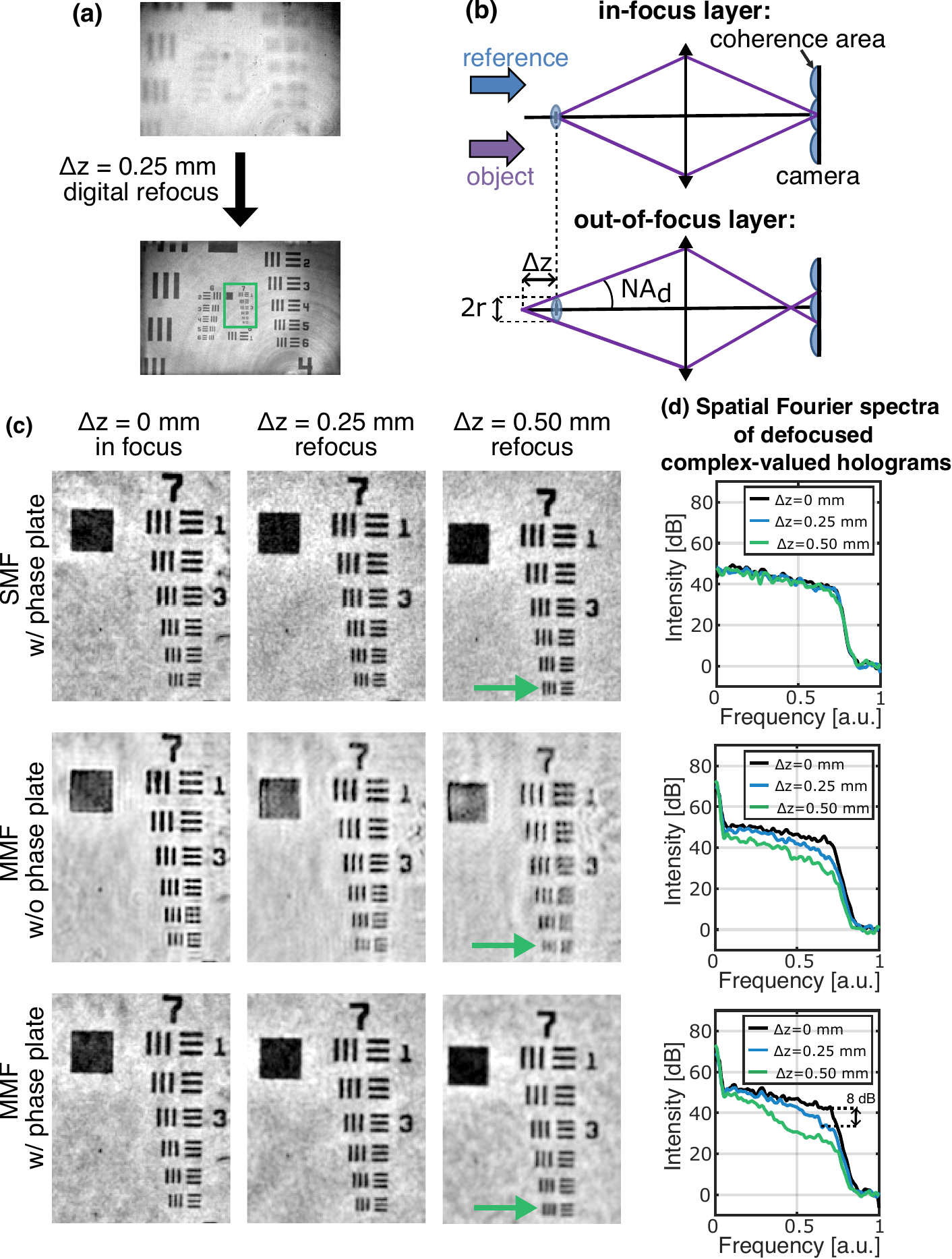}
\caption{Detection NA and achievable axial range with partially spatially incoherent light.
(a) Holographic digital refocusing.
(b) Marginal rays from out-of-focus layers may not impinge on their corresponding coherence areas, resulting in a loss of lateral resolution.
(c) USAF target imaged at different defocuses and digitally refocused. With MMF, lateral resolution deteriorates over depth.
(d) Spatial frequency spectra of complex holograms of paper imaged at different defocuses quantitatively reveals the loss of higher spatial frequencies with the MMF.
}
\label{6_AxialRange_detectionNA}
\end{figure}

Holographic OCT enables layer-wise digital refocusing, which effectively increases the axial range over which OCT can be performed without loss of lateral resolution~\cite{Hillmann2011}. Although aberration correction remains possible when using partially spatially incoherent light~\cite{Borycki2020}, holographic refocusing capabilities at greater depths are hindered~\cite{Auksorius2022_iScience}. We here investigate the respective effects of the detection and illumination NA, and determine the axial range over which refocusing can be performed with acceptable losses.


Digital refocusing of a USAF target imaged with rotating phase plate (100 averaged volumes) defocused by $\Delta z = \SI{0.25}{\milli\metre}$ is illustrated in Fig.~\ref{6_AxialRange_detectionNA}(a). As shown further by looking more closely in the region of interest, the refocused image appears sharp but may not be as sharp as it would have been if the target had been focused on the camera.
This stems from the way coherence areas formed by the reference beam filter defocused light from the sample. As illustrated in Fig.~\ref{6_AxialRange_detectionNA}(b), only light rays from the object impinging on the camera within their corresponding coherence areas can interfere and be detected. For layers in-focus or just slightly defocused, light can be detected and refocused digitally. Light from layers with greater defocuses may however impinge the camera outside their corresponding coherence areas. Rays with larger detection angles -which carry information about higher spatial frequencies- are more rapidly defocused and therefore more rapidly lost. Thus, if the defocus is too large, these marginal rays will not be detected, resulting in a loss of high spatial frequencies, i.e., of lateral resolution.
The problem can be approached with geometrical optics. We consider a defocused object whose ray of largest NA impinges the camera on the edge of the reference field's coherence area, i.e., the limit of detectability. Since the coherence areas in the camera and object planes are conjugate, this ray is also passing by the edge of the coherence area in the object plane. Therefore, the largest defocus $\Delta z$, for which rays corresponding to the detection NA still impinge on the camera within their corresponding coherence area of radius $r$ on sample, can be approximated as $r = {\rm NA_{d}} \Delta z $.
This defines the largest defocus for which diffraction-limited resolution can be achieved with the numerically refocused holograms. For OCT layers beyond this range, light rays with the largest angles permissible within the detection NA should interfere less efficiently, leading to a deterioration of lateral resolution. With coherence areas of diameter $\SI{80}{\micro\metre}$ and ${\rm NA_{d}}=0.174$, the acceptable defocus is $\Delta z=\SI{220}{\micro\metre}$, which gives a total axial range of $\SI{440}{\micro\metre}$ with both sides of the focus position. Since this loss of high frequencies comes from the detection NA, which is the same with and without diffuse illumination, we hypothesized that the phase plate would not change the situation.

This effect is verified experimentally in Fig.~\ref{6_AxialRange_detectionNA}(c) by imaging a USAF target at different defocuses. The sample was placed at a defocus of 0, 0.25, and $\SI{0.50}{\milli\metre}$. The target was imaged with SMF and phase plate, and with MMF with and without phase plate. Holograms were interpolated by a factor of two and digitally refocused by angular spectrum propagation. Images are cropped to show the 7th group of the test target. The width of a line for the 6th element is $\SI{2.2}{\micro\metre}$, the calculated pixel size on sample before interpolation is $\SI{2.4}{\micro\metre}$. Thanks to the suppression of coherence-noise, the lines are clearly resolved at the focus only when using the phase plate.
With SMF and phase plate, the sample can be brought back into focus with the smallest features resolved for all defocuses. With MMF and without phase plate, coherence-noise becomes seemingly stronger over depth, which degrades the resolution. With MMF and phase plate, the features are resolved for a defocus of $\SI{0.25}{\milli\metre}$, but not so clearly for a defocus of $\SI{0.50}{\milli\metre}$. This stands in good agreement with the $z=\SI{220}{\micro\metre}$ acceptable defocus value expected from geometrical optics.
The loss of higher spatial frequencies over depth can be measured quantitatively from the spatial spectrum of the complex-valued holograms. A diffuse sample should however be used for that purpose. Otherwise, for a reflecting sample such as USAF target, the frequency content is dominated by the frequency content of the illumination field, and the spectra for the plane wave and diffuse illuminations would not be comparable.
Therefore, we imaged paper in the same conditions as the USAF target, and the obtained spectra are shown in Fig.~\ref{6_AxialRange_detectionNA}(d). With SMF, the frequency content remains identical at different defocuses. A loss of higher frequencies is however visible with MMF. For the $\SI{0.25}{\milli\metre}$ defocus, a loss of $\SI{8}{\decibel}$ is measured for the highest frequencies with and without diffuse illumination. For the $\SI{0.50}{\milli\metre}$ defocus, the loss of intensity for the highest frequencies is more pronounced, but low frequencies are also affected. These two observations stand in good agreement with our prediction.
The loss of highest frequencies with phase plate is slightly more important than without phase plate for the $\SI{0.50}{\milli\metre}$ defocus, which may be due to optical misalignment.


\begin{figure}[t!]
\centering
\includegraphics[width = 1\linewidth]{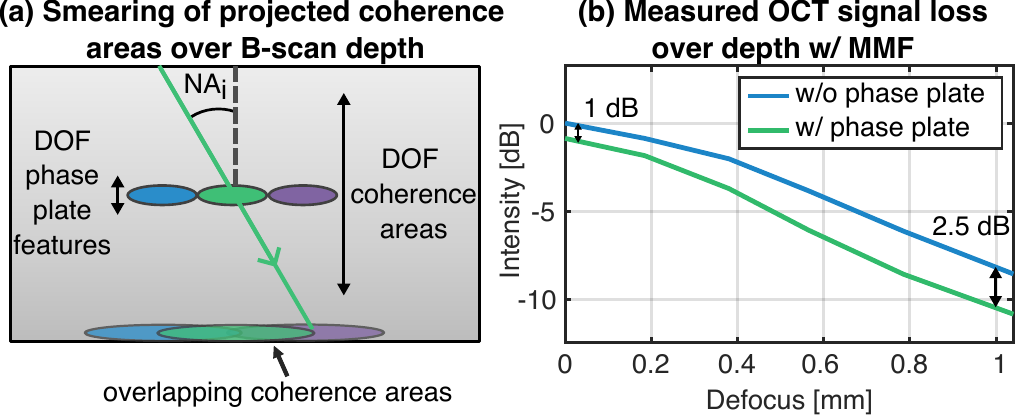}
\caption{Illumination NA and achievable axial range with partially spatially incoherent light.
(a) The high illumination NA causes layers at greater depth to be illuminated with blurred coherence areas.
(b) With MMF, a drop of $\SI{2.5}{\decibel}$ is measured for a $\SI{1}{\milli\metre}$ defocus with diffusion illumination compared to without diffuse illumination.
}
\label{7_AxialRange_illuminationNA}
\end{figure}

With partially spatially incoherent light, the high illumination NA may also reduce the axial imaging range by illuminating layers at depth with blurred coherence areas. If overlapping coherence areas are projected onto the sample and re-imaged on the camera, the mismatch with the coherence areas from the reference beam will prevent light from interfering efficiently. Using the phase plate in combination to the MMF should therefore result in a loss of signal-to-noise ratio over depth.
Without phase plate, Auksorius et al. used the Rayleigh length to estimate the axial range over which the coherence areas are not blurred~\cite{Auksorius2022_iScience}. Calculated as $z_{\rm R} = \pi r^2 / \lambda_0$, with $r = \SI{40}{\micro\metre}$ is the radius of coherence areas, the Rayleigh range is here worth $z_{\rm R} = \SI{5.9}{\milli\metre}$. However, when diffusing the illumination, the Rayleigh length cannot be used to estimate the axial range of coherence areas. As illustrated in Fig.~\ref{7_AxialRange_illuminationNA}(a), the increased illumination NA causes coherence areas to overlap faster when propagating along the optical axis. The depth of focus of the phase plate features, calculated with the formula ${\lambda_0}/{\rm NA^{2}_{i}}$ is $\SI{28}{\micro\metre}$, which corresponds to a few axial pixels. This very short depth of focus for the phase plate features has for consequence that layers are illuminated with different patterns. The depth of focus of coherence areas on the other hand determines at which depth they start overlapping.
We investigated experimentally the loss of signal induced over depth by the diffuse illumination. We imaged paper at different defocuses with MMF and with one half of the field of view diffusely illuminated, and the other half illuminated with a plane wave. The sample was moved backwards out-of-focus while the reference mirror was kept at the same position. The intensity of the OCT signal with and without diffuse illumination measured at different defocuses is shown in Fig.~\ref{7_AxialRange_illuminationNA}(b). Over depth, there is a reduced signal due to the limited temporal coherence of the laser, and because of the attenuation of higher-frequencies of the digital Fourier transform. An additional loss of intensity exists over defocus with the MMF due to the higher spatial frequencies not being detected (as investigated in Fig.~\ref{6_AxialRange_detectionNA}). When diffusing the partially spatially incoherent illumination, a small drop of signal ($\SI{1}{\decibel}$ at $\SI{0}{\milli\metre}$ defocus) that slightly increases over depth is visible. The additional loss due to the diffuse illumination is however very modest, as even for a defocus as large as $\SI{1}{\milli\metre}$, the loss compared to using the MMF without diffuse illumination is only of $\SI{2.5}{\decibel}$. Therefore, the signal loss caused by the illumination NA appears less limiting than the one brought by the detection NA.


\section{Discussion \& conclusion}


We investigated the potential of diffuse illumination and incoherent averaging to improve holographic OCT. In principle, this approach is equivalent to modulating the illumination with a spatial light modulator on the sample arm only and incoherently averaging successive volumes~\cite{Borycki2019STOC, Borycki2020}. In addition, we also studied how diffuse illumination can be combined with a reduction of spatial coherence of the illumination, and investigated whether spatial resolution is increased by the diffuse illumination.
We found that although the contrast in individual volumes is deteriorated by the illumination pattern, averaging multiple volumes with varying illuminations results in largely improved images. Speckle is in practice usually reduced by spatial smoothing, which deteriorate resolution. A reduction of speckle therefore comes as an improvement of the ability to reveal small features of low-reflectivity that would otherwise be lost in the smoothing operation. We found that it is important to differentiate speckle from singly and multiply scattered light. Experimental work and simulations support the notion that speckle from singly scattered light can only be modestly reduced with angular compounding and limited illumination angles~\cite{Winetraub2019, Desjardins2007, Stremplewski2019}.
In our experiments, we found that speckle from multiply scattered light can on the other hand be seemingly arbitrarily reduced. We found that the number of independent speckles for multiply scattered light was large enough to satisfyingly suppress its speckle, although this may depend on the scattering properties of the sample. Since the number of possible optical paths for multiply scattered light is very large, it should be expected that the number of independent speckles obtained by changing the illumination angle be greater for multiply scattered light than for singly scattered light.
This result is consistent with the literature insofar as multiply scattered light is known to decorrelate faster than singly scattered light when varying the detection angle~\cite{Hillman2010}, thereby allowing a higher number of independent speckles to be obtained through the available aperture. In turbid media, multiple scattering light also decorrelates faster than single scattering light~\cite{Thrane2017}. This angular decorrelation of multiply scattered light has been used in scanning OCT to reduce multiple scattering by using a deformable mirror to induce slight changes in the illumination point-spread function~\cite{LiuAdie2018, WuAdie2021}. A similar scheme was used to reduce single scattering speckle~\cite{ZhangZawadzki2019}. One limitation of this method, that also applies to the work of Liba et al.~\cite{Liba2017}, is that a compromise is made between lateral resolution and speckle reduction~\cite{Winetraub2019}. This is because a more strongly aberrated point-spread function worsens lateral resolution but improves speckle decorrelation. In holographic OCT, the diffuse illumination may worsen the contrast of individual volumes but does not reduce the lateral resolution. Holographic OCT is therefore not subject to this tradeoff between spatial resolution and speckle reduction.
Liba et al. claimed they could obtain an unlimited number of independent speckles with their diffuse illumination but did not discuss multiply scattered light. It is worth noting that they used a configuration where the sample is illuminated with diffused light that has an NA possibly greater than the detection NA, which offers a higher number of independent speckles and therefore a better speckle reduction for singly scattered light. As discussed further, increasing the illumination NA  up to the maximum limit imposed by the objective is also possible in holographic OCT.


Diffuse illumination in holographic OCT also brings about notable improvements in image quality by making the detected intensity insensitive to local phase differences. The resulting images offer the distinct advantages of incoherent imaging. The incoherent averaging of volumes obtained with changing diffuse illumination satisfying $\rm{NA_{i}} \geq \rm{NA_{d}}$ suppresses noises arising from spatial coherence, in the same way as fully spatially incoherent illumination does. Important advantages of a coherent detection are however preserved, including the interferometric sensitivity gain and the holographic ability to correct for defocus and higher order aberrations. Temporal phase variations can also still be measured when keeping the diffuser stable. This places diffuse illumination holographic OCT in an advantageous position, where it profits from features of both coherent and incoherent imaging techniques.
In other contexts, increasing the illumination NA is used to improve lateral resolution. Abbe showed in his experiments with coherent light, that multiple slits unresolved with on-axis illumination could be resolved with oblique illumination~\cite{Kohler1981}. From the perspective of the two-point resolution experiment, tilting the illumination essentially changes the phase difference between the slits so that their point-spread function favorably interfere and they can be resolved. Diffusing the light that illuminates the sample from all possible oblique angles within the available NA.
The effect of a diffuse illumination can be considered from a Fourier optics perspective as well. The spectrum of the electric field backscattered by the object is the convolution product of the spectra of the illumination field and of the object, filtered by the detection aperture. A diffuse illumination downshifts higher spatial frequencies from the sample into the passband of the optical system~\cite{WickerHeintzmann2014}.
The gain in resolution is however only obtained when performing a processing akin to deconvolution to shift the higher spatial frequencies back at their correct position. This type of approach can improve by a factor of two the highest resolvable spatial frequencies and have been explored in digital holography~\cite{Cotte2013}, as well as in OCT~\cite{Chowdhury2013, Grebenyuk2018}. Since we do not perform such operation, the resolution of the system in the sense of the sample's highest spatial frequency that can be detected remains unchanged.


We used partially spatially incoherent light to reduce multiple scattering. The confocality stemming from the partial spatial incoherence however hinders the holographic refocusing capabilities. Defocus is problematic enough that the sample must be carefully conjugated with both the camera and the phase plate to maximize signal.
If the defocus is too strong, the interferometric pinholes may reject part of the defocused light rays by preventing them from interfering. Since the defocused signal with greater detection NA spreads over larger surfaces on the camera, higher spatial frequencies are lost first. Loss of signal occurs when the defocused signal extends over an area greater than the size of the coherence areas of the reference beam. Therefore, there is a compromise to reach between the ability to reject multiply scattered light (size of coherence areas), the lateral resolution (detection NA), and the depth range. Interestingly, such compromise had been formulated similarly with Fourier optics for Fresnel reconstructions with partially spatially incoherent digital holography~\cite{DuboisRequena2004}. We formulated this tradeoff with geometrical optics and found good agreement between predicted and experimental values. A detection NA of 0.174 and coherence areas estimated to $\SI{80}{\micro\metre}$ allowed a full axial range of $\SI{0.5}{\milli\metre}$ over which diffraction limited imaging could be achieved. Within this axial range, the diffusion of the illumination did not further decrease the image quality. 
However, diffusing the partially spatially incoherent illumination blurs the coherence areas illuminating the sample more rapidly over depth, leading to a loss of signal-to-noise ratio. We found experimentally that in the situation where ${\rm NA_{i}} = {\rm NA_{d}}$, the loss of sensitivity within the range over which diffraction limited resolution is achieved is nonetheless negligible. Therefore, the constraints stemming from the illumination NA seem not as strong as the one stemming from the detection NA.
This conclusion is important as both speckle- and spatial coherence-noise suppression can be made more effective by increasing the illumination NA.


The phase plate looks overall promising for ophthalmic applications.
As the diffuse illumination reduces speckle from multiply scattered light, structural images of the choroid should be largely improved.
In the presence of diffusing opacities in the anterior part of the eye, retinal images should also be greatly improved thanks to the speckle-reduction of multiply scattered light.
The reduction of speckle-noise of singly scattered light should also improve the imaging of transparent structures like the cornea or the upper retinal layers. Improvement in the retina is however restricted because the natural eye's pupil limits the illumination NA.
More importantly, diffuse illumination could help a potential clinical translation of holographic OCT regarding safety standards. In holographic OCT, the laser is focused close to the eye. Imaging wide fields of view requires bringing the laser focus in the anterior segment, thereby placing the iris at risk of incidental exposure.
Eye safety standards are however generally restrictive for the maximum permissible exposure of the anterior eye. Prominently, ISO 15004-2 sets a limit for group 1 instruments operating at $\SI{840}{\nano\metre}$ to $\SI{1.3}{\watt\per\centi\meter\squared}$ for the posterior part of the eye and only $\SI{20}{\milli\watt\per\centi\meter\squared}$ for the anterior part. As shown in Fig.~\ref{1_Setup_Streuwellen_OAMZ}, the diffuse illumination extends the illumination spot size in the anterior segment by almost 3 orders of magnitude compared to the MMF, and would therefore allow a much more powerful illumination of the eye fundus.
It should be noted that in comparison with a Gaussian distribution, the top-hat distribution spreads the energy in an optimal way for eye safety considerations and should also allow a better decorrelation of the speckle pattern for singly scattered light.
The main downside we foresee to the use of diffuse illumination for ophthalmology is that phase imaging would be compromised because of the motion of the eye with respect to the illumination pattern. Volume-to-volume phase variations introduced by pixels being illuminated with different phases would complicate the detection of local temporal phase changes intrinsic to the retina. Contrasts derived from phase fluctuations such as OCT-angiography, dynamic-OCT, and optoretinography would likely be hindered.

In conclusion, we used a slowly changing diffuse illumination and incoherent averaging of multiple volumes to improve holographic OCT images. Although speckle from singly scattered light can only be modestly reduced with angular compounding, we discovered that speckle arising from multiply scattered light can be on the other hand arbitrarily reduced. Images with largely improved ability to reveal low-reflectivity features are obtained. In contrast to scanning systems, there is no tradeoff between speckle reduction efficiency and spatial resolution. We also showed that the diffuse illumination improves OCT images by removing artifacts arising from spatial coherence, resulting in OCT images seemingly incoherently contrasted. Finally, we combined the diffuse illumination with a MMF to reduce the spatial coherence of the light source and further mitigate multiple scattering. At greater depths, lateral resolution and signal-to-noise ratio are deteriorated in relation with the detection and illumination NA, as well as the size of coherence areas. Holographic capabilities can however be retained over a depth sufficient to many applications with suitable parameters. Overall, diffuse illumination holographic OCT occupies a favorable position, as it profits from advantages of both coherent and incoherent imaging techniques.

\section*{Funding}
\begin{center}
Deutsche Forschungsgemeinschaft (HU 629/6-1).
\end{center}

\section*{Disclosures}
\begin{center}
DB: Zeiss (E, P).
\end{center}

\section*{SUPPLEMENT}
\begin{center}
\textcolor{blue}{\href{https://youtu.be/i6heBqrCOtc}{YouTube video presentation}}
\end{center}

\bibliographystyle{unsrt}
\bibliography{./Bibliography}

\end{document}